\documentclass[twocolumn,printnumbers,amsmath,amssymb,showpacs,prl]{revtex4}
\usepackage{graphicx}
\usepackage{color}

\begin{document}
\title{Mechanical properties of jammed packings of frictionless spheres under applied shear stress}

\author{Hao Liu}
\author{Hua Tong}
\author{Ning Xu$^*$}

\affiliation{CAS Key Laboratory of Soft Matter Chemistry, Hefei National Laboratory for Physical Sciences at the Microscale, and Department of Physics, University of Science and Technology of China, Hefei 230026, People's Republic of China.}

\begin{abstract}
By minimizing a thermodynamic-like potential, we unbiasedly sample the potential energy landscape of soft and frictionless spheres under constant shear stress.  We obtain zero-temperature jammed states under desired shear stresses and investigate their mechanical properties as a function of the shear stress.  As a comparison, we also obtain jammed states from the quasistatic-shear sampling in which the shear stress is not well-controlled.  Although the yield stresses determined by both samplings show the same power-law scaling with the compression from point $J$, i.e.~the jamming transition point at zero temperature and shear stress, for finite size systems, the quasistatic-shear sampling leads to a lower yield stress and a higher critical volume fraction of point $J$.  The shear modulus of jammed solids decreases when increasing the shear stress.  However, the shear modulus does not decay to zero at yielding.  This discontinuous change of the shear modulus implies the discontinuous nature of the unjamming transition under nonzero shear stress, which is further verified by the observation of a discontinuous jump of the pressure from jammed solids to shear flows.  The pressure jump decreases upon decompression and approaches zero at the critical-like point $J$, in analogy with well-known phase transitions under external field.  The analysis of force networks in jammed solids reveals that the force distribution is more sensitive to the increase of the shear stress near point $J$.  The force network anisotropy increases with the shear stress.  Weak particle contacts near the average force and under large shear stresses exhibit asymmetric angle distribution.
\end{abstract}


\maketitle

\section{Introduction}
Athermal particulate systems such as foams and granular materials undergo the jamming transition denoted as point $J$ when the packing density increases to a critical value$^{[1-5]}$.  This jamming transition is signified by the sudden formation of rigidity to resist shear and compression, which resembles typical noncrystalline liquid-solid transitions such as the glass transition and the yielding of amorphous solids under shear.  Compared to the glass transition and yielding, the jamming transition at point $J$ seems simpler, due to the absence of dynamics.  People believe that there are common underlying mechanisms for these different transitions, although they are driven by different parameters, e.g. volume fraction (density) $\phi$ for the jamming transition and glass transition, temperature $T$ for the glass transition, and shear stress $\sigma$ for the yielding, which have inspired the idea of unifying these different transitions in the same framework of the jamming phase diagram$^{[1,2,6]}$.

The standing of point $J$ in the jamming phase diagram is significant. It is the upper limit of the glass transition of hard spheres and the ending point of the yielding of jammed packings of repulsive particles in the zero shear stress limit.  Therefore, studying the jamming transition at point $J$ and properties of jammed solids at $T=0$ will definitely shed light on our understanding of noncrystalline liquid-solid transitions in general.  As a unique transition, the jamming transition at point $J$ exhibits unusual critical behaviors, which have received extensive concerns in the past decade$^{[7-24]}$.  Most of previous studies have been focused on marginally jammed solids near point $J$ at $T=0$ and $\sigma=0$.  Only recently, the influence of applied shear stress on jamming has attracted some attention.  For instance, it has been shown that the critical volume fraction $\phi_c$ at point $J$ moves to a higher value when the jamming threshold is probed by quasistatically sheared systems$^{[8,25]}$.

In addition to the jamming transition, how jammed solids respond to applied shear stress is of great importance to understanding the mechanical failure of materials, especially granular materials.  As is well known, in granular materials, particle contacts constitute a force network$^{[26]}$.  Due to the structural disorder, contact forces are spatially heterogeneous and show a wide distribution$^{[27-30]}$.  It has been proposed that the force network can be separated into two subnetworks, a strong one with contact forces higher than the average and a weak one otherwise.  When being sheared, these two subnetworks exhibit both geometric and mechanical anisotropy which are interestingly complementary$^{[31-33]}$.  However, a concrete picture of the evolution of the force network anisotropy with applied shear stress is still lacking.

The major difficulty to study jamming under applied shear stress is the lack of an efficient method to explore the potential energy landscape and find local potential energy minima under the constraint of constant shear stress.  In most of previous simulations, people applied quasistatic shear by successively applying a step shear strain followed by the energy minimization$^{[25,34-38]}$.  In such an approach, the shear stress is not a well-controlled quantity, while in reality shearing is mostly triggered by the shear force instead of the shear strain.  Recently, we reported an efficient method to find local potential energy minima under constant shear stress by minimizing a thermodynamic-like potential$^{[7]}$.  This new method enables us to generate jammed solids under desired shear stress and thus study how their properties vary with the shear stress by unbiasedly sampling the potential energy landscape.

As briefly discussed in Ref.~[7], compared to the unbiased sampling of our new method, the quasistatic shear tends to explore low-energy regions of the potential energy landscape.  It is then interesting to know whether and how these two approaches differ from each other in the description of jamming under applied shear stress.  We use CS and QS to denote the constant shear stress sampling and quasistatic-shear sampling.  In this paper, we are mainly concerned about the mechanical properties of jammed solids under applied shear stress, including the yield stress, shear modulus, and force networks, with comparisons between the CS and QS samplings.  The structure of the paper is as follows.  In Section 2, we introduce the details of our simulation model and the two samplings.  In Section 3, we compare the QS sampling with the CS sampling in the measurement of the critical scaling of the yield stress.  In Section 4, we discuss the discontinuous nature of the unjamming transition at $\sigma>0$ by showing discontinuous change of the shear modulus and pressure.  In Section 5, we are focused on the shear stress dependence of the force distribution and force network anisotropy.  We conclude our work in Section 6.


\section{Simulation model and methods}
Our systems are three-dimensional boxes with side length $L$ in all directions, which contain $N$ frictionless spheres.  We apply Lees-Edwards boundary condition in the $z$ direction and periodic boundary conditions in both $x$ and $y$ directions to mimic shearing$^{[39]}$.  The shear force is applied in the $x$ direction and the shear gradient is in the $z$ direction.  To avoid crystallization, we use $N/2$ large and $N/2$ small particles with equal mass $m$ and a diameter ratio $1.4$.  The interaction potential between particles $i$ and $j$ is harmonic:
\begin{equation}
U_{ij}=\frac{\epsilon}{2}\left(1-\frac{r_{ij}}{d_{ij}}\right)^2\Theta\left(1-\frac{r_{ij}}{d_{ij}}\right),
\end{equation}
where $r_{ij}$ is their separation, $d_{ij}$ is the sum of their radii, and $\Theta(x)$ is the Heaviside function.  We set the units of mass, energy, and length to be $m$, $\epsilon$, and small particle diameter $d_s$.

To perform the QS sampling, we start with $1000$ distinct jammed states generated by minimizing the total potential energy $U=U(\vec{r}_1,\vec{r}_2,...,\vec{r}_N,\gamma=0)$ of random states using the fast inertial relaxation engine (FIRE) minimization method$^{[40]}$.  A step shear strain $\Delta\gamma = 10^{-4}$ is then applied to these states, followed by the FIRE minimization of $U(\vec{r}_1,\vec{r}_2,...,\vec{r}_N,\gamma=\Delta\gamma)$.  The same procedure is repeated until the system is sheared up to a unit strain $\gamma=1$.  The whole process therefore mimics the shear flow in the zero shear rate limit.  For each initial state, we obtain $10000$ states under different shear stresses.  As shown in the inset to Fig.~1(a), every point on the $\sigma-\gamma$ curve represents a state sampled by the QS sampling.  Apparently, in the QS sampling, states sampled are historically dependent and the shear stress cannot be controlled.  To investigate the shear stress dependence of the states obtained by the QS sampling, we just bin the shear stress using an interval $\Delta$ and average states with the shear stress lying in $(\sigma-\Delta/2, \sigma + \Delta/2)$.

In order to obtain jammed states under desired shear stress $\sigma$, we start with random states and minimize the thermodynamic-like potential$^{[7]}$
\begin{equation}
H(\vec{r}_1,\vec{r}_2,...,\vec{r}_N,\gamma)=U(\vec{r}_1,\vec{r}_2,...,\vec{r}_N,\gamma)-\sigma\gamma L^3, \label{H}
\end{equation}
using FIRE minimization method.  Eq.~(\ref{H}) is the Legendre transformation of the potential energy $U(\vec{r}_1,\vec{r}_2,...,\vec{r}_N,\gamma)$ to change the control parameter from the shear strain $\gamma$ (QS sampling) to the shear stress $\sigma$ (CS sampling).  By minimizing $H$, local potential energy minima with a shear stress $\sigma$ can be obtained, so we are able to realize the CS sampling.  Different from the QS sampling in which the energy minimization is performed under a fixed shear strain, the shear strain in the minimization of the CS sampling is changing.  The shear strain is initially set to be zero, which increases during the minimization until a jammed state with the shear stress $\sigma$ is found.  Since the initial states are randomly selected, this CS sampling unbiasedly samples the potential energy landscape.

\section{Critical scaling of the yield stress}

In Ref.~[7], we apply the CS sampling to measure the probability of finding jammed states under constant shear stress.  At $\phi>\phi_c$, when the shear stress is small, the probability is one, while it decays to zero when the shear stress is large.  We define the yield stress as the shear stress at which the probability is $0.5$.  It is thus the critical shear stress in the potential energy landscape perspective, above which it is hard to find jammed states able to sustain applied shear stress.  The yield stress of jammed systems with harmonic repulsion from the CS sampling is linearly scaled with $\phi-\phi_c$.

\begin{figure}
\includegraphics[width=0.45\textwidth]{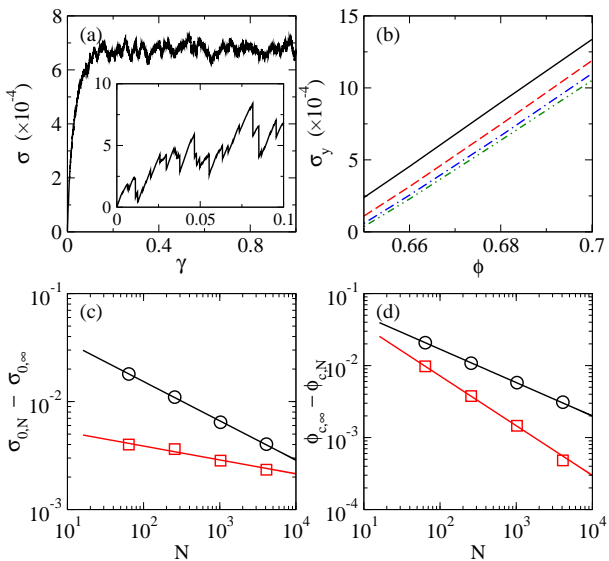}
\caption{Shear stress $\sigma$ versus the shear strain $\gamma$ for jammed solids consisting of $N=1024$ particles at $\phi=0.68$ averaged over $1000$ independent runs of quasistatic shear.  The inset shows an example of a single QS run.  (b) Volume fraction $\phi$ dependence of the yield stress $\sigma_y$ for systems with $N=64$, $256$, $1024$, and $4096$ from the top to the bottom.  (c) and (d) show the system size dependance of the fitting parameters $\sigma_{0,N}$ and $\phi_{c,N}$ for the CS (circles) and QS (squares) samplings.  The lines are the fits to the data using the formula discussed in the text.  \label{fig:fig1}}
\end{figure}

In literature, the yield stress is usually defined in the shear flow perspective, i.e. it is the average shear stress in the zero shear rate limit.  It can be obtained by fitting the flow curve $\sigma(\dot\gamma)$ with empirical expressions$^{[41]}$, e.g. Herschel-Buckley relation $\sigma=\sigma_y+A\dot\gamma^{\alpha}$, where $\dot\gamma$ is the shear rate, $\sigma_y=\sigma(0)$ is the yield stress, and $A$ and $\alpha$ are fitting parameters.  This yield stress can also be directly measured from the QS sampling.  Figure~1(a) shows $\sigma(\gamma)$ averaged over $1000$ independent QS runs.  A single QS flow consists of lots of segments of elastic deformations (linear increase of $\sigma$ with $\gamma$) followed by a plastic event (sudden stress drop), as shown in the inset to Fig.~1(a).  In average, the shear stress exhibits an initial increase with $\gamma$ before reaching a plateau.  The yield stress is thus defined as the plateau value of the shear stress.

Figure~1(b) shows the volume fraction and system size dependence of the yield stress measured from the QS sampling.  The yield stress is linear in the volume fraction well above point $J$: $\sigma_y=\sigma_{0,N}\left( \phi-\phi_{c,N}\right)$, where $\sigma_{0,N}$ and $\phi_{c,N}$ are the fitting parameters.  $\phi_{c,N}$ is also the average volume fraction of point $J$ for finite systems with $N$ particles.  In Figs.~1(c) and 1(d), we compare the system size dependence of $\sigma_{0,N}$ and $\phi_{c,N}$ between the QS and CS samplings.  For both samplings, we fit $\sigma_{0,N}$ and $\phi_{c,N}$ into $\sigma_{0,N}=\sigma_{0,\infty}+a_{\sigma}N^{-b_{\sigma}}$ and $\phi_{c,N}=\phi_{c,\infty}-a_{\phi}N^{-b_{\phi}}$ with values of the fitting parameters shown in Table~1.

\begin{table}
\centering
\resizebox{8cm}{!}{
\begin{tabular}{cccccccccccccccccccccccccccccc}
\hline\hline
&&  && \vline && $\phi_{c,\infty}$ && \vline && $a_{\phi}$ && \vline && $b_{\phi}$ && \vline && $\sigma_{0,\infty}$ && \vline && $a_{\sigma}$ && \vline && $b_{\sigma}$ && \\
\hline
&& CS && \vline && 0.649 && \vline && 0.137 && \vline && 0.457 && \vline && 0.018 && \vline && 0.082 && \vline && 0.363 && \\
&& QS && \vline && 0.649 && \vline && 0.173 && \vline && 0.691 && \vline && 0.018 && \vline && 0.007 && \vline && 0.128 &&\\
\hline\hline 
\end{tabular}
}
\caption{\label{table:exponents}  Fitting parameters of the power-law scaling of $\phi_{c,\infty} - \phi_{c,N}$ and $\sigma_{0,N} - \sigma_{0, \infty}$ shown in Figs. 1(c) and 1(d).}
\end{table}

For finite size systems studied here, the yield stress from the CS sampling is larger than the QS sampling.  The gap between the two yield stresses decreases when the system size increases.  The yield stress from the CS sampling is by definition the maximum stress under which jammed states can survive, while the yield stress from the QS sampling is the average stress of jammed states under shear.  For finite size systems, the yield stress from the CS sampling measures the upper limit of the stress fluctuations as shown in the inset to Fig.~1(a), which explains the existence of the gap between the two yield stresses.  When system size increases, the stress fluctuations decay, which lead to the decrease of the gap.  It is thus expected that in the thermodynamic limit the two yield stresses are equal because stress fluctuations vanish, as claimed as well in Ref.~[42].

Figure~1(d) indicates that $\phi_{c,N}$ estimated from the QS sampling is larger than the CS sampling.  In Ref.~[7], we attribute it to the fact that the QS sampling tends to explore low-energy states, which is a demonstration of the historic dependence of the QS sampling.  For finite size systems, the volume fraction of the jamming transition at point $J$ is not well-defined, but broadened to a distribution with a width $w$ due to the finite size effect$^{[2]}$.  $w$ decreases when increasing the system size and vanishes in the thermodynamic limit $^{[2]}$.  In the distribution, the probability of finding jammed states $P_J(\phi)$ increases from $0$ to $1$ when the volume fraction $\phi$ increases.  For unbiased samplings like the CS sampling, $P(\phi)=\frac{{\rm d}P_J(\phi)}{{\rm d}\phi}$ is approximately Guassian-like, and $\phi_{c,N}$ is the volume fraction at which $P(\phi)$ is the maximum and the probability of finding jammed states is approximately $0.5$.  However, because the QS sampling tends to explore low-energy states, it is more likely to find unjammed states, so the probability of finding jammed states being $0.5$ is pushed to a higher $\phi_{c,N}$ than the CS sampling.  In the thermodynamic limit, $\phi_{c,\infty}$ should be well-defined and independent of ways of sampling, we thus expect that $\phi_{c,\infty}$ are identical for both samplings.  As shown in Fig. 1, we use the same $\phi_{c,\infty}$ and $\sigma_{0,\infty}$ for both the CS and QS samplings and can indeed fit the data well.

\section{Discontinuous unjamming transition at $\sigma>0$}

When the applied shear stress increases to the yield stress, marginally jammed solids are expected to be softened and unjam to shear flows.  As discussed above, in the thermodynamic limit, the yield stress vanishes at the well-defined point $J$ at $\phi_{c,\infty}$.  Criticality of point $J$ has been widely discussed$^{[7-24]}$, although the nature of the jamming transition at point $J$ has not come to the conclusion.  If we assume that point $J$ is critical, the jamming phase diagram in the $\sigma-\phi$ plane at $T=0$ reminisces typical phase transitions like the paramagnetic-ferromagnetic and superconducting transitions under external magnetic field, in which the transition is critical in the absence of the magnetic field while discontinuous when the field is nonzero.  Is the unjamming transition at $\sigma>0$ also discontinuous?

\begin{figure}
\includegraphics[width=0.45\textwidth]{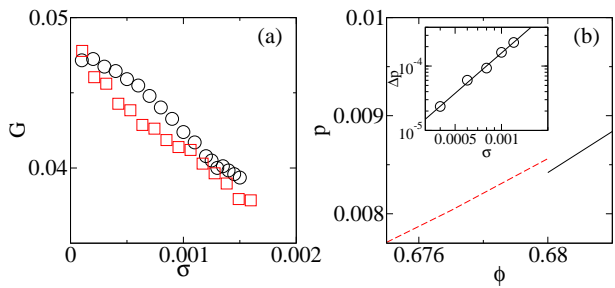}
\caption{(a) Shear stress $\sigma$ dependance of the shear modulus $G$ measured from the CS (circles) and QS (squares) samplings. The systems are jammed solids consisting of $N=1024$ particles at $\phi=0.70$.  (b) Volume fraction dependence of the pressure measured at $\sigma=10^{-3}$ for $N=1024$ systems.  Solid and dashed curves are measured for the jammed solids and unjammed shear flows, respectively.  The inset shows the shear stress dependence of the pressure gap between unjammed and jammed states, $\Delta p$, at yielding for $N=1024$ systems. The line shows the relation $\Delta p\sim \sigma^2$. \label{fig:fig2}}
\end{figure}

In Fig.~2(a), we compare the shear moduli $G$ of jammed solids obtained from the CS and QS samplings under different shear stresses up to the yield stress.  For both samplings, the shear modulus decreases and the solid is softened when the shear stress increases.  Under the same shear stress, the shear modulus from the QS sampling is in general a little bit smaller than the CS sampling.  This is consistent with the fact that the QS sampling moves $\phi_{c,N}$ to a higher value than the CS sampling.  Because the shear modulus is proportional to $(\phi-\phi_{c,N})^{1/2}$ $^{[2]}$, at the same volume fraction, the jammed solids from the QS sampling are closer to point $J$ and thus have a smaller $G$.  Figure~2(a) also shows that for both samplings the shear modulus does not go to zero at yielding.  When the applied shear stress is higher than the yield stress, the system flows for ever with $G=0$, so there is a discontinuous change of the shear modulus from $G>0$ to $G=0$ at yielding when the yield stress $\sigma_y>0$.  This provides an evidence that the unjamming transition at $\sigma>0$ is discontinuous.

Figure~2(b) shows another evidence of the discontinuous unjamming transition at $\sigma>0$.  Under a fixed shear stress $\sigma$, when the volume fraction increases, systems undergo the jamming transition from unjammed (steadily flowing) states to jammed solids at the crossover volume fraction $\phi_y$ determined by $\sigma=\sigma_y(\phi_y)$.  When $\phi<\phi_y$, the CS sampling can also generate unjammed states under the constant shear stress $\sigma$.  We can then compare the pressure of jammed solids with that of unjammed flowing states at $\phi_y$.

As shown in Fig.~2(b), there is a small discontinuity in the pressure around $\phi_y$.  The pressure of unjammed states is higher than that of jammed solids at $\phi_y$.  Note that here we only take into account the excess part of the pressure arising from particle interactions and exclude the ideal gas contribution, because our CS minimization is not a real dynamical process and does not capture particle velocities.  For unjammed states, if the shear rate is zero at the yield stress, the ideal gas term of the pressure is negligible.  However, previous studies have shown that, for viscoelastic shear flows driven by constant shear stress, the critical shear rate at the yield stress defined as the minimum shear stress below which systems eventually cease flowing may not vanish for finite size systems$^{[42,43]}$.  If then, there will be an ideal gas contribution to the pressure of unjammed states at the yield stress, which leads to an even larger pressure gap than that shown in Fig.~2(b).  In any case, the presence of the pressure gap, although small compared to the pressure itself, implies as well that the unjamming transition at $\sigma>0$ is discontinuous.

The inset to Fig.~2(b) shows the pressure gap, $\Delta p=p_u-p_j$, at yielding as a function of the shear stress for $N=1024$ systems, where $p_u$ and $p_j$ are pressures of unjammed and jammed states.  Our data suggest a possible relation $\Delta p\sim \sigma^2\sim p^2\sim G^4$, which indicates the trend of decreasing the gap with decreasing the shear stress.  This is consistent with the criticality of point $J$ in the thermodynamic limit and in analog with those conventional phase transitions under external magnetic field. Our study here presents some evidences to suggest the discontinuous nature of the yielding (unajamming transition at $\sigma>0$) of jammed states as a possible phase transition.  However, further investigations are required to verify our suggestion with the key to find the correct order parameter to characterize the transition.

\section{Anisotropic force networks in jammed solids at $\sigma>0$}

For jammed packings of grains, e.g.~sand piles, contact forces between grains form a spatially heterogeneous force network.  When subject to a load, the force network becomes anisotropic and directional, forming percolating force chains to support the load$^{[32]}$.  Jammed packings of spheres near point $J$ are typical model systems to study force networks in granular materials.  It has been shown that the distribution of the contact force $P(F)$ in jammed packings of frictionless spheres near point $J$ in general decreases when increasing the force with an exponential decay at large $F$$^{[27-30,33]}$.  For systems with harmonic repulsion, a maximum in $P(F)$ at small $F$ emerges when the jamming transition is approached from the unjammed side$^{[27]}$.  By applying the CS sampling to generate jammed solids under desired shear stresses, we are able to explore statistically how the force distribution and force network anisotropy evolve with the shear stress.

\begin{figure}
\includegraphics[width=0.45\textwidth]{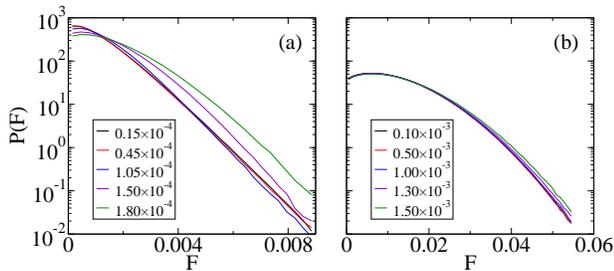}
\caption{Contact force distributions $P(F)$ of jammed solids consisting of $N=1024$ particles at (a) $\phi=0.65$ and (b) $\phi=0.70$ and different shear stresses labeled in the legends.   \label{fig:fig3}}
\end{figure}

Figure~3 shows the evolution of the force distribution with the shear stress at $\phi=0.65$ and $0.70$, averaged over $10000$ states for each stress.  At both volume fractions, there is a maximum in $P(F)$ at small $F$.  However, the maximum is more pronounced and shifts to a higher value of $F/\left< F\right>$ at $\phi=0.70$ than at $\phi=0.65$, with $\left< F\right>$ the average contact force.  Moreover, the large $F$ tail of $P(F)$ at $\phi=0.65$ is exponential, while it decays faster than exponential at $\phi=0.70$ as already reported before$^{[27]}$.  The more remarkable difference between the two volume fractions is their distinct response to the increase of the applied shear stress.  At $\phi=0.65$, when the shear stress approaches the yield stress, the maximum at small $F$ is lowered and broadened, and $P(F)$ increases significantly at large $F$. These indicate that there are more large force contacts under larger shear stress.  At $\phi=0.70$, however, $P(F)$ is almost constant under all the shear stresses, so the number of large force contacts does not change apparently, unlike at $\phi=0.65$.  The much stronger response of $P(F)$ to the applied shear stress at $\phi=0.65$ than at $\phi=0.70$ evidences the growth of the susceptibility, if the susceptibility is defined as the response of $P(F)$ to the applied shear stress.  The susceptibility is directly linked to the spatial correlation, assuming that the fluctuation-dissipation theorem is satisfied, so the spatial correlation is expected to grow approaching point $J$, implying again the criticality of Point J.  To verify this picture, more thorough and intensive future work is needed to quantify the volume fraction dependence of the susceptibility and find out if it exhibits any critical scaling.

\begin{figure}
\includegraphics[width=0.45\textwidth]{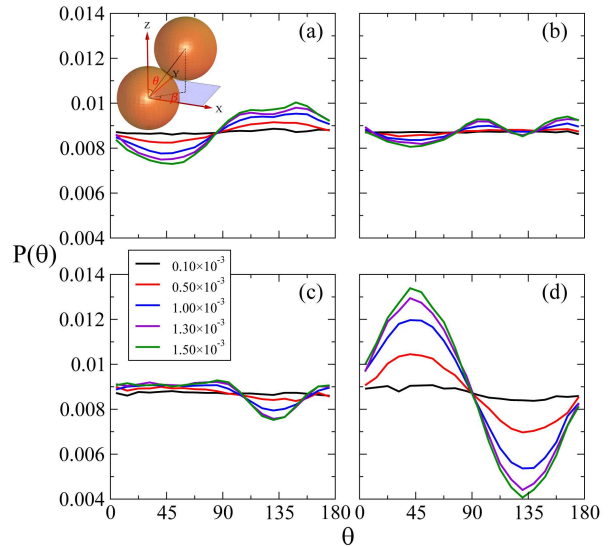}
\caption{Distribution of the angle $\theta$ of particle contacts, $P(\theta)$, measured in four force windows: (a) [0, 0.005], (b) [0.005, 0.0115], (c) [0.0115, 0.015], and (d)[0.030, 0.040], for jammed solids consisting of $N=1024$ particles and at $\phi=0.70$.  The average force is $\left< F\right>=0.0115$.  The five curves in each panel are measured under different shear stresses with the values shown in the legend.  The inset to (a) illustrates the definition of the angles $\theta$ and $\beta$.  \label{fig:fig4}}
\end{figure}

As suggested by previous studies$^{[31-33]}$, the average contact force $\left< F\right>$ separates the force network of a jammed solid into two subnetworks, a strong ($F>\left< F\right>$) and a weak ($F<\left< F\right>$) ones.  The strong network is load bearing and percolating, while the weak one is dissipative and does not support the load.  When subject to a planar shear with the shear direction in the $x$ direction and shear gradient in the $z$ direction, it is expected that the strong network consists of force chains aligned in the $x-z$ plane to resist the shear.  To verify that and see how the force network anisotropy evolve with the shear stress, we show in Fig.~4 the distribution of the angle $\theta$ defined in the inset to Fig. 4(a).  For each contact, we are concerned about two angles $\theta$ and $\beta$ in the spherical coordinates, as illustrated in the inset to Fig.~4(a).  In Fig.~4, we are focused on the distribution of $\theta$, $P(\theta)$, satisfying
$\int_{0^\circ}^{180^\circ}P(\theta)\sin(\theta)\,\mathrm{d}\theta=1$, for contacts lying in $\beta\in [135^\circ, 225^\circ]$, which are mostly affected by the shear.

Near point $J$, the average force changes apparently with the shear stress, as shown in Fig. 3, so the self-averaging disappears $^{[27]}$.  Moreover, fluctuations become larger approaching point $J$. Both effects lead to uncertainties in the quantification of the force network anisotropy.  In Fig. 4, we demonstrate the evolution of the force network anisotropy at $\phi=0.70$ because the force distribution is almost independent of the shear stress, so that we can be focused on the anisotropy without being disturbed by other changes.

We divide the whole force network into many subnetworks.  In each subnetwork, the forces are within the same force window.  In Fig. 4, we show the evolution of $P(\theta)$ with the shear stress in four force windows.  Panels (a) and (b) are below the average force $\left< F\right>\approx 0.0115$, while panels (c) and (d) are above.  Interestingly, $\left< F\right>\approx 0.0115$ acts as a boardline.  The general shape of $P(\theta)$ at $F>\left< F\right>$ is symmetric to that at $F<\left< F\right>$ with respect to $\theta=90^\circ$.  This validates the use of $\left< F\right>$ to separate strong and weak contacts.  For the weak subnetworks with $F\ll\left< F\right>$, there is a minimum in $P(\theta)$ at $\theta\approx 45^\circ$, which decreases when increasing the shear stress, while the majority of the weak contacts point to a $\theta$ lying in $(90^\circ, 180^\circ)$.  For the strong subnetworks with $F\gg\left< F\right>$, the majority of the contacts point to $\theta\approx 45^\circ$, indicating that the strong contacts tend to form force chains perpendicular to weak contacts.  One may expect that the weak contacts mostly point to $\theta\approx 135^\circ$, which is approximately true with a peak in $P(\theta)$ at $\theta\approx 135^\circ$ when the forces are much smaller than $\left< F\right>$ and the shear stress is small, as shown in Fig.~4(a).  However, when $F$ is close to $\left< F\right>$ and the shear stress is large, the peak at $\theta=135^\circ$ splits into two peaks at around $100^\circ$ and $170^\circ$, as shown in Fig.~4(b).  Such a split has not been observed before, which is more pronounced with the increase of the shear stress.  When the forces are slightly larger than $\left< F\right>$, as shown in Fig.~4(c), the effect of the split still exists and leads to an asymmetric $P(\theta)$ as well under large shear stresses.  The occurrence of the split and its dependence on the shear stress is not trivial, because it is robust near $\left< F\right>$ and under large shear stresses, which deserves further investigations.  When the forces are much larger than $\left< F\right>$, Fig. 4(d) indicates that those forces are more and more aligned in the principle direction at $\theta=45^\circ$ when increasing the shear stress, which should be the main supporters of the shear.

\section{Conclusions}

Following Ref.~[7], we study mechanical properties of jammed solids under constant shear stress and compare the results from CS and QS samplings.  All our analyses are benefited from the CS sampling which can unbiasedly explore the potential energy landscape under constant shear stress and search for jammed solids under desired shear stress.  Although most of the analyses can be performed using the QS sampling, we have demonstrated that the QS sampling is biased and cannot accurately control the shear stress and is thus awkward in the study of shear stress dependence.

For finite size systems, the QS sampling leads to a smaller yield stress and a larger average critical volume fraction of the jamming transition at point $J$ ($T=0$ and $\sigma=0$).  In the thermodynamic limit, our analysis suggests that the differences between the two samplings vanish.  By statistically measuring the shear modulus as a function of the shear stress, we find that the shear modulus decreases with the increase of the shear stress.  However, the shear modulus does not decay to zero at yielding, implying the discontinuous nature of the unjamming transition at $\sigma>0$.  The discontinuity is further verified by the discontinuous jump of the pressure at yielding from jammed solids to unjammed shear flows.  The pressure jump decreases when decreasing the shear stress and shows the trend to vanish at point $J$ when $\sigma=0$, which is analogous to typical phase transitions under external field.  Moreover, we investigate the evolution of the force distribution and force network anisotropy.  The force distribution shows weak shear stress dependence away from point $J$, while it is sensitive to the change of the shear stress near point $J$, again implying the criticality of point $J$.  When subject to a shear stress, strong contact forces are aligned in the direction of $\theta=45^\circ$ to support the shear, while weak forces distribute around $\theta=135^\circ$.  Interestingly, when the forces are close to the average force and the shear stress is large, the peak in the distribution $P(\theta)$ at $\theta=135^\circ$ splits into two peaks, which leads to an asymmetric $P(\theta)$.  The underlying physics of this split requires further investigations.

\section{Acknowledgement}

This work is supported by National Natural Science Foundation of China (Grant No. 21325418), National Basic Research Program of China (973 Program, Grant No. 2012CB821500), CAS 100-Talent Program (Grant No. 2030020004), and Fundamental Research Funds for the Central Universities (Grant No. 2340000034).

\end{document}